\documentclass[aps,prl,twocolumn,superscriptaddress,showpacs]{revtex4}
\usepackage{graphicx}
\usepackage{color}
\begin{document}

\title{Reversal of Nonlocal Vortex Motion in the Regime of Strong Nonequilibrium}


\author{Florian Otto}
\altaffiliation{Present address: atto{\bf cube} systems AG, Germany.}
\affiliation{Institute for Experimental and Appl. Physics, University of Regensburg, D-93025 Regensburg, Germany}
\author{Ante Bilu\v{s}i\'{c}}
\affiliation{Institute for Experimental and Appl. Physics, University of Regensburg, D-93025 Regensburg, Germany}
\affiliation{Faculty of Natural Sciences, University of Split, N. Tesle 12, HR-21000 Split, Croatia}
\author{Dinko Babi\'{c}}
\affiliation{Department of Physics, Faculty of Science, University
of Zagreb, Bijeni\v{c}ka 32, HR-10000 Zagreb, Croatia}
\author{Denis Yu.~Vodolazov}
\affiliation{Institute for Physics of Microstructures, Russian
Academy of Sciences, 603950, Nizhny Novgorod, GSP-105, Russia}
\author{Christoph S\"{u}rgers}
\affiliation{Physikalishes Institut and Center for Functional Nanostructures Universit\"{a}t Karlsruhe, D-76128 Karlsruhe, Germany}
\author{Christoph Strunk}
\affiliation{Institute for Experimental and Appl. Physics, University of Regensburg, D-93025 Regensburg, Germany}

\date{\today}

\begin{abstract}
We investigate nonlocal vortex motion in weakly pinning $a$-NbGe nanostructures, which is driven by
a transport current $I$ and remotely detected as a nonlocal voltage $V_{nl}$. At high $I$, the
measured $V_{nl}$ exhibits dramatic sign reversals that at low and high temperatures $T$ occur for
opposite polarities of $I$. The sign of $V_{nl}$ becomes {\it independent} of that of the
drive current at large $|I|$. These unusual effects can be nearly quantitatively explained by a
novel enhancement of magnetization, arising from a nonequilibrium distribution of quasiparticles at
high $T$, and a Nernst-like effect resulting from local electron heating at low $T$.

\end{abstract}

\pacs{74.25.Qt,74.25.Fy,74.78.Db,74.78.Na}
\maketitle

Motion of the Abrikosov vortex lattice in type-II superconductors results in strong deviations
of the quasi-particle distribution function from that in equilibrium
\cite{Larkin2,Kunchur,Kunchur2} when the lattice is strongly driven by a transport current. Close
to the critical temperature $T_c$, overheating of quasiparticles within the vortex cores leads to a
{\it shrinkage} of the cores, accompanied by decreasing the effective viscosity coefficient $\eta$ - the
Larkin-Ovchinnikov (LO) instability \cite{Larkin2}, while the quasiparticles outside the cores remain in thermal equilibrium. At low $T$, the entire quasiparticle subsystem is heated because of the larger electron-phonon collision time. This results in an {\it expansion} of the cores instead of their shrinkage, while $\eta$ again decreases
\cite{Kunchur,Kunchur2}. In both cases, the current-voltage [$V(I)$] characteristics are very nonlinear - can become even hysteretic \cite{BabicNbGe} - and are in fact so similar that the difference can be resolved only via a quantitative analysis  \cite{BabicNbGe,Babic}. However, vortex shrinkage and vortex expansion are different effects and should lead to qualitative differences in other properties.

\begin{figure}
\includegraphics[width=75mm]{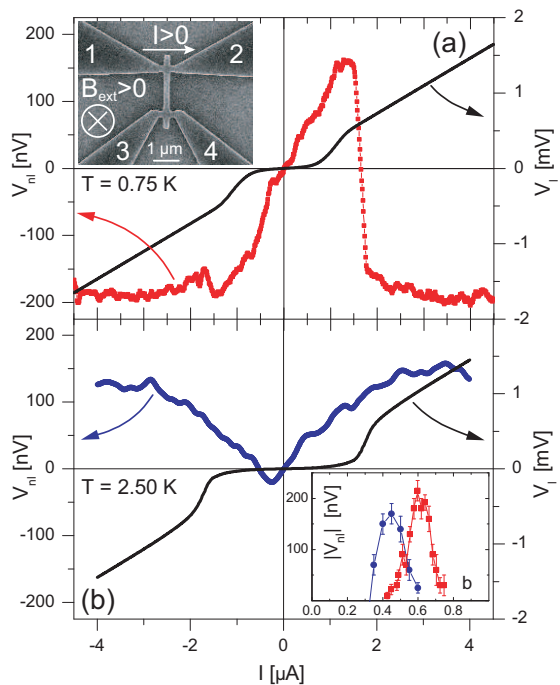}
\caption{\label{figure1}Typical local (black lines) and nonlocal (colored symbols)
$V(I)$ curves. (a) $T=0.75\,\mathrm{K}=0.26\,T_c$ (at $B_{ext}=3.0\,\mathrm{T}$, $b=0.64$),
and (b) $T=2.50\,\mathrm{K}=0.85\,T_c$ (at $B_{ext}=0.50\,\mathrm{T}$, $b=0.50$).
Inset to (a): Sample geometry; $L=2\,\mathrm{\mu m}$, $W=250\,\mathrm{nm}$.
Inset to (b): Saturation voltages of $V_{nl}$ plotted against $b$, for low (red)
and high (blue) $T$.}
\end{figure}

In this Letter, we report novel effect in the recently discovered nonlocal vortex flow in the {\it transversal}
flux transformer geometry (TFTE) \cite{Grigorieva,Helzel}, which allow a clear distinction of the above two opposite
types of nonequilibrium. We apply a drive current $I$ in one part of the sample (local lead) and
measure the voltage response ($V_{nl}$) in a remote part of the superconductor connected with first
one via a channel of the same material  [see the inset to Fig.~\ref{figure1}(a)]. In such a geometry, one
can probe changes in the vortex lattice which occur in the local lead via changes in the
interaction between vortices in the local lead and vortices in the rest of the sample. In this way,
we can detect a novel nonequilibrium enhancement of the magnetization of the superconductor in the
LO state with respect to the equilibrium magnetization and observe a Nernst-like signal at low
$T$.

Previously, $V_{nl}(I)$ was investigated in the linear response regime \cite{Grigorieva,Helzel}.
The main features of these studies can be accounted for by a simple model of locally driven
vortices pressurizing those in the channel by repulsive vortex-vortex interaction \cite{Helzel}.
$I$ applied between the contacts 1 and 2 in the inset to Fig.~\ref{figure1}(a) decreases
exponentially in the perpendicular channel, with a decay length $W/\pi \ll L$
\cite{Grigorieva,OttoDiss}. Thus, $\approx n_\phi WX$ driven vortices face $\approx n_\phi WL$
vortices in the channel, where $X$ is the effective length over which the driving force $f_{dr}$
(per unit vortex length $d$) acts, $n_\phi=B / \phi_0$ the vortex density, $\phi_0$ the magnetic
flux quantum, $B = B_{ext} + \mu_0 M$, $B_{ext} $ the external magnetic field, $M$ the magnetization, and $\mu_0 = 4 \pi \cdot
10^{-7}\,\mathrm{Vs/Am}$. The driven vortices push or pull those in the channel by exerting a
pressure $p=(n_\phi W X) (f_{dr} / W)$. The resulting force $pWd$ is balanced by the total
frictional force $(n_\phi WL ) (\eta u_{nl}d)$ on the vortices in the channel (which move at
velocity $u_{nl}$). For a superconductor with a large magnetic penetration depth $\lambda$,
i.e., $n_\phi \approx B_{ext} / \phi_0$, using $V_{nl} = W B_{ext} u_{nl}$ for the voltage detected
at the probes 3 and 4, one obtains
\begin{equation}\label{equation1}
V_{nl} =  WB_{ext}X f_{dr} / \eta L \quad.
\end{equation}
At low $I$, i.e., close to equilibrium, $f_{dr}$ is given by the Lorentz force $f_L= j\phi_0$,
where $j$ is the transport current density. In Ref.~\cite{Helzel}, $X=W$ led to
$V_{nl}=(WB_{ext}\phi_0/\eta L d) I = R_{nl} I$. This reproduced the observed $V_{nl} \propto I$
and $V_{nl} \propto 1/L$ even in the presence of pinning \footnote{Pinning affects $R_{nl}$, but could be accounted for by a modified $\eta$.}.

Our $d=40$ nm thick $a$-Nb$_{0.7}$Ge$_{0.3}$ samples are produced by electron-beam lithography and magnetron sputtering. The local current leads (1,2) are connected to the nonlocal voltage probes (3,4) via a perpendicular channel of $L=2$ $\mu$m and
$W=250$ nm. All data for $V_l(I)$ refer to passing $I$ between 1 and 3, and measuring $V_l$ between
2 and 4. Since $W$ is also the width of all other narrow sample parts, in particular that linking 1
and 2, $V_l(I)$ and $V_{nl}(I)$ can be compared directly. Measurements of $V_l(I)$ provided all
relevant parameters of our samples: $T_c=2.94\,\mathrm{K}$, the normal-state resistivity
$\rho_n=1.82\,\mu\Omega\mathrm{m}$, $-(dB_{c2}/dT)_{T=T_c}=2.3\,\mathrm{T/K}$, where $B_{c2}$ is
the equilibrium upper critical magnetic field, and the Ginzburg-Landau (GL) parameters $\kappa=72$,
$\xi(0)=7.0\,\mathrm{nm}$, and $\lambda(0)=825\,\mathrm{nm}$. The low pinning in
$a$-Nb$_{0.7}$Ge$_{0.3}$ allowed for dc measurements of $V_{nl} \sim 10 - 200\,\mathrm{nV}$, which
was at the level of $R_{nl} \sim 0.1$ $\Omega$ in the low-$I$ linear regime. All measurements were
carried out in a $^3$He cryostat, with $B_{ext}$ perpendicular to the film plane.

Typical results for the two limiting cases of low
($T=0.75\,\mathrm{K} = 0.26\,T_c$) and high ($T=2.50\,\mathrm{K} =
0.85\,T_c$) temperatures are shown in Fig.~\ref{figure1}(a) and
Fig.~\ref{figure1}(b), respectively.
The $V_l(I)$ curves exhibit a
nonlinear shape characteristic of strong-nonequilibrium (SNEQ),
originating either in (a) electron heating \cite{Kunchur,Babic,BabicNbGe} or (b)
LO vortex-core shrinking \cite{Larkin2,Babic,BabicNbGe}. On the
other hand, $V_{nl}(I)$ displays the previously observed linear,
antisymmetric dependence [i.e., $V_{nl}(-I) = - V_{nl}(I)$] only at
low $I$. Upon increasing $I$, sudden sign reversals of $V_{nl}$ are
observed in both regimes: at a certain $I$, the antisymmetric signal
converts into a symmetric one. The sign of $V_{nl}$ can be
unambiguously attributed to the following directions in the inset to
Fig.~\ref{figure1}(a): at low positive (negative) $I$, the positive
(negative) $V_{nl}$ corresponds to vortex motion upwards (downwards)
in the channel. When $I$ is high, vortices move either
downwards ($T \ll T_c$, $V_{nl}<0$), or upwards ($T \rightarrow T_c $ ,
$V_{nl} >0$),  {\it irrespective of the direction of} $I$. The saturation values of
$| V_{nl} |$ at high $I$ are plotted vs $b=B_{ext}/B_{c2}$ in the inset
to Fig.~\ref{figure1}(b). In both cases, nonzero values are observed
only at intermediate $b$, with a maximum efficiency around $b=0.6$
($b=0.45$) at low (high) $T$, similarly to the previously observed
$B_{ext}$ sweep traces of $V_{nl}$ at low $I$ \cite{Grigorieva,Helzel}.
As argued in Ref.~\cite{Helzel}, the vanishing of $V_{nl}$
at low $B_{ext}$ is presumably related to $n_\phi$
becoming smaller than the density of pinning
sites, whereas $V_{nl}(B_{ext}
\rightarrow B_{c2}) \rightarrow 0$ because the sample goes to the normal state.

We first discuss the regime $T\ll T_c$.
Assigning the corresponding high-$j$ SNEQ state to
electron heating to $T=T^*$ above the bath temperature $T_0$ was
successful in explaining the measured $V_l(I)$ of
Refs.~\cite{Kunchur,Babic,BabicNbGe}.
An analysis of the present $V_l(I)$ \cite{OttoDiss}
within the same framework
permits to extract $T^*(V_l)$ and, using $V_l(I)$,
also $T^*(I)$, which is more convenient for a comparison
with the $V_{nl}(I)$ data (see below).
The hot electrons penetrate
into the channel, which remains at $T=T_0$, roughly up to
$L_T=\sqrt{D \tau_0} \approx 295\ \mathrm{nm}\sim W$. Here, $D=4.80\cdot
10^{-5}\,\mathrm{m^2/s}$ is the diffusion constant, and
$\tau_0\approx 1.82 \,\mathrm{ns}$ is the relaxation time of the hot
electrons, resulting from the mentioned analysis \cite{OttoDiss}.
Hence, there is a $T$ gradient which leads to a thermal driving force
${\bf f}_T= - S_\phi \nabla T$ and consequently to the Nernst effect.
$S_\phi$ is the vortex transport entropy \cite{HuebenerBook}.
The Nernst effect should lead to vortex motion downwards, which agrees with the observed $V_{nl} <0$.
Since $T^* - T_0 \sim 1$ K typically, the observed temperature gradients $| \nabla T |
\sim (T^*-T_0)/L_T \sim 1$ K/$\mu$m are much
larger than in usual measurements of the Nernst effect.

\begin{figure}
\includegraphics[width=70mm]{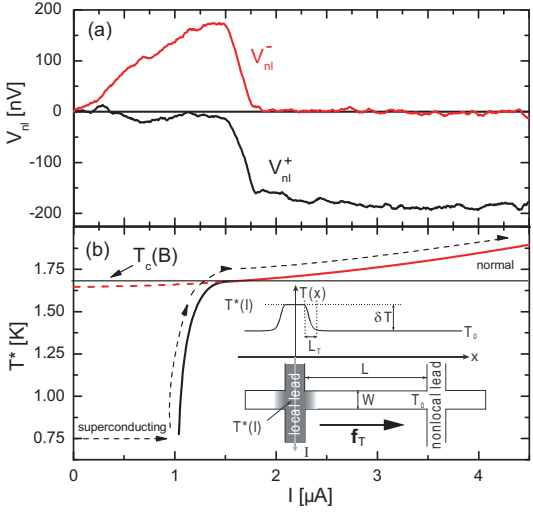}
\caption{\label{figure2}(a) Measured $V_{nl}$ at $T\ll T_c$: antisymmetric (red)
and symmetric (black) part of the nonlocal signal at
$T=0.75\,\mathrm{K}$ and $B_{ext}=3.0\,\mathrm{T}$.
(b) Effective electron temperature $T^*(I)$, where the black line stems from an analysis of $V_l(I)$, and the red line from noise measurements in the normal state. Inset:
Sketch of the temperature profile along the channel.}
\end{figure}

The above is elaborated in Fig.~\ref{figure2}, where the result for
$B_{ext}=3.0\,\mathrm{T}$ ($b=0.64$) is analyzed more closely. The
shape of $V_{nl}(I)$ in Fig.~\ref{figure1}(a) suggests to consider the symmetric (+) and
antisymmetric ($-$) parts of $V_{nl}$ separately via
$V_{nl}^{\pm}(I)=[V_{nl}(I) \pm V_{nl}(-I)]/2$, which is shown in
Fig.~\ref{figure2}(a). $V_{nl}^{-}(I)$ at low $I$ is fairly linear as
expected, since $f_{dr} = f_L$, while $V_{nl}^{+}(I)$ is very small. Upon increasing $I$, this is followed by a rapid
suppression of $V_{nl}^{-}(I)$ and a simultaneous growth of
$V_{nl}^{+}(I) < 0$ to a constant value comparable to that of the
maximum $V_{nl}^{-}(I)> 0$. Returning to Fig.~\ref{figure1}(a), one can note
that this dramatic change occurs around $I$ where $V_l(I) \approx R_n I$, signifying the transition to the normal state in the local region \cite{Babic,BabicNbGe}
and consequent vanishing of $f_L$.
Furthermore, $| I |$ where the sign of $V_{nl}$ changes steeply on the $I > 0$ side
(${\bf f}_L$ and ${\bf f}_T$ act oppositely)  coincides
with $| I |$ where  $V_{nl}$ has a local minimum on the $I < 0$ side (${\bf f}_L$ and ${\bf f}_T$ add); in both cases, this marks that
only ${\bf f}_T$ remains effective at higher $| I |$.

In the main panel of Fig.~\ref{figure2}(b), we plot $T^*(I)$ extracted according to
the electron heating model \cite{Babic,BabicNbGe} in the
superconducting state and from noise measurements in the normal
state \cite{OttoDiss}, whereas in the inset we show
a sketch of the $T$ profile along the sample.
One can see that the electron heating is basically absent
at low $I$, then sets in very steeply until it reaches
$T_c(B_{ext})$ that represents $B_{c2}(T)$ \cite{Babic,BabicNbGe,OttoDiss},
after which it changes with $I$ only weakly.
The nearly flat $V_{nl}^{+}(I)$ at high $I$ hence corresponds to $T^* \approx
T_c(B_{ext})$, so
$| \nabla T | \approx [T_c(B_{ext})-T_0] / L_T=\delta T / L_T$.
Using Eq.~(\ref{equation1}), we
can extract $S_\phi$ from our data by focusing on the saturating
values of $V_{nl}^{+}(I)$. We approximate
$f_{dr}=f_T \approx S_\phi \delta T /L_T$
and $X \approx L_T$ to obtain
\begin{equation}\label{equation2}
S_\phi = V_{nl} \phi_0 / R_{nl} \delta T d \quad ,
\end{equation}
which does not contain $L_T$.
Since $S_\phi$ and $R_{nl}$ depend on the properties of the
channel (where $T=T_0$), the observed
$V_{nl}^{+}(I) \approx \mathrm{const.}$ follows straightforwardly.
In the $(B_{ext},T)$ range of our data, we
find $S_\phi  \sim 0.1-1.5 \cdot 10^{-12}$
Jm$^{-1}$K$^{-1}$ \cite{OttoDiss}, which is
in reasonable agreement with a theoretical estimate $\sim 0.1-0.2
\cdot10^{-12}$ Jm$^{-1}$K$^{-1}$ obtained by using the Maki formula
\cite{Maki, Kopnin}, as well as with experimental data
on films of Nb ($0.05-1.5\cdot10^{-12}$
Jm$^{-1}$K$^{-1}$) \cite{Huebener} and of
Pb-In ($0.2-5\cdot10^{-12}$ Jm$^{-1}$K$^{-1}$) \cite{Vidal}.

\begin{figure}
\includegraphics[width=75mm]{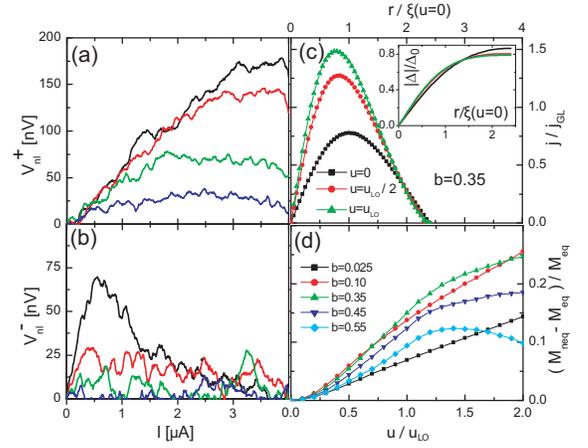}
\caption{\label{figure3}Measured  $V_{nl}^{+}(I)$ (a), and $V_{nl}^{-}(I)$ (b)
at $T=2.50\,\mathrm{K}=0.85\,T_c$ and $b=0.45$ (black), $0.50$
(red), $0.55$ (green) and $0.60$ (blue). (c) Calculated $j_s/j_{GL}$
and $|\Delta|/\Delta_0$ (inset) vs $r/\xi(u=0)$ for different
$u/u_{LO}$. (d) Calculated $(M_{neq}-M_{eq})/M_{eq}$ against
$u/u_{LO}$ at different $b$ (as indicated).}
\end{figure}

We now turn to the regime $T \rightarrow T_c$. An analysis \cite{OttoDiss} of the $V_l(I)$ in the
spirit of Refs.~\cite{Babic,BabicNbGe} reveals that this SNEQ state corresponds to the LO
vortex-core shrinking \cite{Larkin2}, with $T \approx T_0$ everywhere because electron heating is
strongly suppressed close to $T_c$ \cite{Kunchur,Babic,BabicNbGe}. $V_l(I)$ for $T \ll T_c$ and $T
\rightarrow T_c$ are at first glance rather similar, so the difference becomes obvious only
through a numerical analysis \cite{Babic,BabicNbGe}. In contrast, the qualitatively different
$V_{nl}(I)$ curves in Fig. 1 leave no doubt that we are dealing with two distinct SNEQ phenomena. As before,
the shape of $V_{nl}(I)$ [see Fig.~\ref{figure1}(b)] suggests to consider $V_{nl}^{+}(I)$ and
$V_{nl}^{-}(I)$ separately, which is shown in Figs.~\ref{figure3}(a) and \ref{figure3}(b),
respectively. $V_{nl}^{-}(I)$ at low $I$ is linear for small $b$, which implies the presence of
$f_L$, whereas this is difficult to claim for higher $b$ where the signal is small over the entire
$I$ range. At high $I$, however, $V_{nl}^{-}(I)$ is small regardless of $b$. $V_{nl}^{+}(I)$, on
the other hand, increases with increasing $I$, and eventually saturates at a value comparable to
that of the maximum Nernst signal at low $T$, albeit with the opposite sign. The smallness of
$V_{nl}^{-}(I)$ at high $I$ suggests inefficiency of $f_L$ in this regime. This can be understood
by recalling [see Fig.~\ref{figure1}(b)] that the $V_l(I)$ for these $I$ is close to the
normal-state dissipation, which means that most of the current is normal \cite{Larkin2} - and
normal current does not contribute to $f_L$.

Since $f_L$ is negligible and $T \approx T_0$,
there must be yet another driving
force which governs the TFTE at high $I$.
Below we show that this force has the same origin as the LO
effect on $V_l(I)$, that is,
a deviation $\delta g(\epsilon)$ of the quasiparticle
distribution function $g(\epsilon)$ from
$g_{eq}(\epsilon)=\tanh(\epsilon/2k_BT) = g(\epsilon) - \delta g(\epsilon)$
in equilibrium. An additional consequence of $\delta g$ is
an enhancement of the supercurrent
density $j_s$ flowing around the vortex core, which can be
calculated following \cite{Larkin2,Schmid}
\begin{equation}
{\bf j}_s=\frac{1}{\rho_n
e}\left(\frac{\pi}{4k_BT_c}|\Delta|^2+\frac{\pi}{2} |\Delta|\delta
g(|\Delta|)\right)\left(\nabla \varphi-\frac{2e}{\hbar}{\bf
A}\right) \, ,
\end{equation}
where $\Delta=|\Delta|\mathrm{exp}(i\varphi)$ is the order
parameter and {\bf A} the vector potential. The term $\propto|\Delta|^2$ corresponds to the
equilibrium contribution to ${\bf j}_s$ in the GL model, and the
term $\propto \delta g$ to the SNEQ correction. $\delta g$ is
positive for energies less than the maximal value $| \Delta |_{max}$ of
the order parameter in a single-vortex cell \cite{Larkin2}, and
$|\Delta|$ is enhanced near the vortex core
[see the inset to Fig.~\ref{figure3}(c)]. Both these factors lead to a growth of
${\bf j}_s$ near the vortex core [see Eq.~(3)]. Therefore, the
magnetic moment ${\bf m} = (1/2) \int [{\bf r} \times {\bf j}_s]dS_{cell}$
of each cell in the vortex lattice increases in the LO state.

We base our model on addressing ${\bf m}$ to find $M$ along the direction of $B_{ext}$,
which is alternative (but much simpler with regard to the role of $\delta g$)
to using the Gibbs free energy density for the same purpose.
Qualitatively, ${\bf m}$ of a given
cell creates a dipole magnetic field which in the surrounding cells
opposes $B_{ext}$, hence an increase of $j_s$ results in a stronger diamagnetic response.
Note that the same argument can be used to explain increase of the equilibrium
diamagnetism of the mixed state as $T$ decreases.
Quantitatively, we have to determine $g(\epsilon)$ and $|\Delta|$. We follow the
LO model and solve numerically the modified GL
equation for $|\Delta|$ (see Eq. (A49) in
\cite{Larkin2}) coupled with the equation for $g(\epsilon)$ (see Eq. (A45) in \cite{Larkin2}).

In Fig.~\ref{figure3}(c), we plot exemplary $j_s /j_{GL}$ vs reduced radial coordinate $r /\xi$, where
$j_{GL} \simeq 0.93\Delta_0 (1-T/T_c)^{1/2} / \xi \rho_n e$ is the equilibrium GL depairing
current density,
$\Delta_0 \simeq 3.06 k_B T_c(1-T/T_c)^{1/2}$, and $\xi$ corresponds to that at
zero vortex velocity $u$.
Results are shown for three different $u$
relative to the LO vortex velocity $u_{LO}$ \cite{Larkin2}; the corresponding
$| \Delta | /  \Delta_0$ is shown in the inset by the same colors.
By summing up the resulting ${\bf m}$ of each cell, one can find
the difference $\delta M  = M_{neq}-M_{eq}$ of the nonequilibrium ($M_{neq}$) and
equilibrium ($M_{eq}$) magnetization. This is presented in Fig.~\ref{figure3}(d).
The maximum of $\delta M / M_{eq}$
occurs for $b \sim 0.2$ (at $u/u_{LO} \approx 1$). At smaller $b$, the
enhancement of $j_s$ near the core gives a small contribution to ${\bf m}$.
At larger $b$, the suppression of $| \Delta |$ at the cell boundary [due
to $\delta g(\epsilon)<0$ for $\epsilon >|\Delta|_{max}$] becomes important.
We show results up to $u=2u_{LO}$, where the LO approach becomes invalid
at $T \sim 0.85 \, T_c$.

The spatial variation of $M$ across the boundary
between the local region and the channel occurs over a length
of about the intervortex distance
$a_0 \approx \sqrt{\phi_0/B_{ext}}$, and induces a current density
${\bf j}_M={\bf\nabla}\times{\bf M}$ that flows along that boundary.
This current creates a force $f_M = j_M \phi_0$ that is
again {\it independent} of the direction of $I$, pulls the
vortices toward the local lead (which results in $V_{nl} > 0$),
and dominates the total $f_{dr}$ in
the SNEQ regime near $T_c$.
The typical $| \delta M | \simeq | M_{eq} | \simeq (B_{c2} -B_{ext})/2\mu_0 \kappa^2 \simeq
35\,\mathrm{A/m}$ ($\hat{=}\,88\,\mu$T at
$B_{ext}=0.45\,\mathrm{T}$) is rather small but appears over a very
small distance $a_0(B_{ext}=0.45\,\mathrm{T})\simeq 70\,\mathrm{nm}$,
thus providing $j_M \simeq 500\,\mathrm{MA/m^2}$
which is of the same order as
the transport current densities we used - as $I=1\,\mathrm{\mu A}$
corresponds to $j=100\,\mathrm{MA/m^2}$. We again employ Eq.~(1) to estimate
$V_{nl}$. Since  $j_M=\partial M/\partial x \approx | \delta M | / a_0$
and $X \approx a_0$, with $f_{dr}= f_M$ we obtain
\begin{equation}
V_{nl}=[WB_{ext}a_0/\eta L] j_M \phi_0 = R_{nl} |\delta M| d \; ,
\end{equation}
from which $a_0$ has dropped out again. Inserting typical values of $R_{nl}\approx 0.1\,
\Omega$ and $|\delta M| \approx 35 \,\mathrm{A/m}$, we find
$V_{nl}\approx 140 \,\mathrm{nV}$, which is quite close to the
measured values.

In view of the simplicity of our model, the
agreement between the experiment and theory is rather remarkable.
A full quantitative account of the phenomenon would require inclusion of
other effects on the interface of the local region
and the channel - such as details of entry/exit
trajectories for the fast and slow vortices, etc. However, these corrections may
depend on the sample geometry, and we believe that the main physics of
TFTE close to $T_c$ is captured by our model.

In conclusion, nonlocal measurements allowed us to qualitatively distinguish two different types of
 vortex motion in strong nonequilibrium. According to our theory, close to $T_c$ a new type of
nonequilibrium magnetization is built up in the drive wire, which pulls the vortices {\it towards}
the drive channel. At low temperatures, electron heating leads to a Nernst effect, which pushes
vortices {\it away} from the drive channel. Remarkably, this happens irrespectively of the sign of
the drive current in both cases. The qualitative features as well as the absolute values of the
observed nonlocal voltages agree well with the results of our model calculations. Our results offer
a new possibility to probe the presence of vortices or vortex-like excitations as currently
discussed in the context of cuprate superconductors \cite{kokanovic}.

\begin{acknowledgments}
We acknowledge discussions with I. Kokanovi\'{c}, V. Vinokur, Y.
Galperin and R. Gross, and financial support by the DFG within GK
638. A.B. acknowledges support from the Croatian Science Foundation (NZZ). D.Y.V. acknowledges support from Dynasty Foundation.
\end{acknowledgments}

%

%
\end{document}